\documentclass[fleqn]{article}
\usepackage{amsmath}
\usepackage{espcrc2}
\usepackage{graphicx}
\title{Dynamics of the Scalar Condensate in thermal
4D self-interacting Scalar Field Theory on the Lattice}

\author{P. Cea\address[DF,INFN]{Dipartimento di Fisica, Univ. of Bari and INFN - Sezione di Bari,
        I-70126 Bari, Italy},
        M. Consoli\address[CT]{INFN - Sezione di Catania,  I 95129 Catania, Italy}, and
        L. Cosmai\address[INFN]{INFN - Sezione di Bari, I-70126 Bari, Italy}}

\begin{document}

\begin{abstract}
We simulate a
 four dimensional self-interacting scalar field theory
 on the lattice at finite temperature.
 By varying temperature,
 the system undergoes a phase
 transition from  broken phase to 
 symmetric phase. Our data show that the
 zero-momentum field
 renormalization increases by
 approaching critical temperature.
 On the other hand,
 finite-momentum wave-function
 renormalization remains remarkably
 constant.
\vspace{1pc}
\end{abstract}

\maketitle

Traditionally, the `condensation' of a scalar field, i.e. the transition
from a symmetric phase where $\langle \Phi \rangle=0$
to the physical vacuum where
$\langle \Phi \rangle \neq 0$, has been described as an
essentially classical phenomenon in terms of
a classical potential
\begin{equation}
\label{clpot}
   V_{\text{cl}}(\Phi) =
 \frac{1}{2} r_0 \Phi^2 +
\frac{\lambda_0}{4!} \Phi^4
\end{equation}
with non-trivial absolute minima at $\Phi=\pm v_B \neq 0$
(`B=Bare'). In this picture, the `scalar condensate' is treated as
a classical c-number field which is simply
taken into account through a shift of the scalar
field, say $\Phi(x) = v_B +h(x)$. In this picture, by
retaining terms at most quadratic in the shifted field $h(x)$, one
expects a simple relation
\begin{equation}
\label{Mh}
           M^2_h= V''_{\text{cl}}(\Phi = v_B)
\end{equation}
relating the `Higgs mass' $M_h$ directly
to the quadratic shape of the potential at the minimum.
Beyond the tree-approximation,
and on the basis of perturbation theory, Eq.~(\ref{Mh})
is believed to represent a good approximation provided
the classical potential is replaced with the full quantum
effective potential $V_{\text{eff}}$.

However,  there is~\cite{Consoli:1994jr,Consoli:1999ni} an
alternative non-perturbative description of spontaneous symmetry
breaking in which the zero-momentum limit is not smooth in the
broken phase, reflecting the discontinuity inherent in a
Bose-Einstein condensation phenomenon, where a single mode
acquires a macroscopic occupation number. In this picture, $M^2_h$
and the curvature of the effective potential at the non-trivial
minima are {\it different} physical quantities related by an
infinite renormalization in the continuum limit of quantum field
theory.

To test this prediction objectively one can proceed as follows.
Let the zero-momentum two-point function (the inverse
susceptibility) be related to the Higgs mass through a re-scaling
factor $Z=Z_\varphi$ {\em defined} as
\begin{equation}
\label{chichi}
 \chi^{-1}=  \Gamma_2(p=0)=
\left. \frac{ d^2 V_{\text{eff}} }{d \varphi^2_B}
\right|_{\varphi_B= \pm v_B} \equiv \frac{M^2_h}{Z_\varphi} \,.
\end{equation}
This definition of $Z=Z_\varphi$ is, {\it a priori}, different from
the quantity $Z=Z_{\text{prop}}$ defined from a single-pole form of the
propagator
\begin{equation}
\label{zprop} G_{\text{pole}}(p)= \frac{Z_{\text{prop}}}{p^2 +
m^2_{\text{prop}}}  \,.
\end{equation}
Indeed, the two quantities agree only if Eq.~(\ref{zprop}) remains
valid down to $p_\mu=0$ (up to small perturbative corrections).

In the continuum limit the alternative description
of SSB predicts that $Z_\varphi \to \infty$ but $Z_{\text{prop}}
\to 1$. In this respect, this picture is very different from
renormalized perturbation theory where the zero-momentum limit is
smooth and the continuum limit corresponds to
\begin{equation}
\label{pertu} \left. Z_\varphi \right|_{\text{pert}} = Z_{\text{
prop}}[1 + \mathcal{O}(\lambda_R)]  \,.
\end{equation}
To measure on the lattice $Z_\varphi$ one has to measure
$\Gamma_2(0)$ and $M_h$. While the zero-momentum two-point
function is directly obtained from the inverse susceptibility
$\chi^{-1}$, determining $M_h$ requires to study the lattice
propagator. This will be described below. Our numerical
simulations  were performed (using the Swendsen-Wang cluster
algorithm)  in the Ising limit of the theory where a one-component
$(\lambda\Phi^4)_4$ theory becomes
\begin{equation}
\label{ising}
S = -\kappa \sum_{x,\mu} \left[
\phi(x+\hat{\mu})\phi(x) + \phi(x-\hat{\mu})\phi(x) \right]
\end{equation}
with $\Phi(x)=\sqrt{2\kappa}\phi(x)$ and where $\phi(x)$ takes the
values $\pm 1$. We measured the bare magnetization
\begin{equation}
\label{baremagn} v_B=\langle |\Phi| \rangle \,, \Phi \equiv \sum_x
\Phi(x)/L^4 \,,
\end{equation}
the zero-momentum susceptibility
\begin{equation}
\label{suscep} \chi=L^4 \left[ \left\langle |\Phi|^2 \right\rangle
- \left\langle |\Phi| \right\rangle^2 \right]  \,,
\end{equation}
and the shifted-field propagator
\begin{equation}
\label{shifted} G(p)= \sum_x \exp (ip x) \left\langle h(x) h(0)
\right\rangle \, ,
\end{equation}
$p\equiv( p_4,{\mathbf{p}})$ with ${\mathbf{p}}=2\pi
\mathbf{n}/L_s$ and $p_4=2\pi n_4/L_t$ $(n_1,n_2,n_3,n_4)$ are
integer-valued components, not all zero. Zero-temperature
simulations correspond to $L_t=L_s=L$ and will be discussed first.

To determine the Higgs
boson mass one has to compare the lattice propagator with the
(lattice version of the) single-pole form Eq.~(\ref{zprop}) by
extracting $M_h=m_{\text{prop}}$ from a two-parameter fit to the
lattice data:
\begin{equation}
\label{gbetaform} G^{(\beta)}_{\text{fit}}(n, {\mathbf{ p}} )= { {
\frac{Z_{\text{prop}}}{\hat{{\mathbf{p}}}^2 + \hat { p_4 }^2 +
m^2_{\text{prop}} } }}
\end{equation}
where $m_{\text{prop}}$ is the mass in lattice units and
$\hat{p}_\mu= 2 \sin(p_\mu/2)$.
In the symmetric phase (i.e. for values of the hopping parameter
$\kappa < \kappa_c \sim 0.0748$), Eq.~(\ref{gbetaform}) provides a
very good description of the lattice data for the propagator in
the whole range of Euclidean momenta, namely from $p=0$ up to $p^2
\sim \Lambda^2$ ($\Lambda \sim \frac{\pi}{a}$ is the ultraviolet
cutoff).

In the broken phase, however, the attempt to fit all data with the
same pair of parameters $(m_{\text{prop}}, Z_{\text{prop}})$ gives
unacceptably large values of the normalized chi-squared.  Still,
one can obtain a good-quality fit with Eq.~(\ref{gbetaform}) by
suitably restricting the range of momenta. In this case there are
two choices: (i) a high-momentum range: $p^2_{\text{min}} \leq p^2
\leq \Lambda^2$ (with $p^2_{\text{min}} \neq 0$); (ii) a
low-momentum range: $0\leq p^2 \leq p^2_{\text{max}}$ ( with
$p^2_{\text{max}} \ll \Lambda^2$). If we require agreement of
$Z_{\text{prop}}$ with $Z_{\text{pert}}$ (the normalization of
massive single-particle states as computed in perturbation theory)
the Higgs mass turns out to be determined from the set (i). Indeed
we found ~\cite{Cea:1999kn} that the choice (i) to compute the
Higgs boson mass and the normalization of single-particle states
satisfies several consistency checks. As concern $Z_\varphi$
introduced in Eq.~(\ref{chichi}), our previous
analysis~\cite{Cea:1999kn}  shows that there is a discrepancy
between $Z_\varphi$ and $Z_{\text{prop}}$. Moreover, this
discrepancy becomes larger when $\kappa \to \kappa_c$, where the
purely perturbative corrections $\mathcal{O}(\lambda_R)$ in
Eq.~(\ref{pertu}) vanish. For instance, on a $32^4$ lattice at
$\kappa=0.07512$, a fit to the propagator data gives
~\cite{Cea:1999kn} $M_h=m_{\text{prop}}=0.206(4)$ and
$Z_{\text{prop}}=0.9551(21)$. However, the measured susceptibility
is $\chi=193.1\pm 1.7$, so that $Z_\varphi=2 \kappa \chi
m^2_{\text{prop}} = 1.234(50)$, at more than $5\sigma$'s from
$Z_{\text{prop}}$.

The same discrepancy can be presented in a different way.
To check that the discrepancy is not due to finite-size effects,
we have repeated the measurement of $\chi$
for $\kappa=0.07512$ on a $40^4$ lattice with the result
$\chi=190.9\pm1.6$.
If we now take the mass value $m_R=0.20$ (for
$\kappa=0.0751(1)$) from Table~3 in~\cite{Luscher:1988ek} we get
$Z_\varphi=2 \kappa \chi m^2_R=1.147(10)$ at about 14$\sigma$'s from the
perturbative prediction $Z_{\text{pert}}=0.937(12)$ in the same Table.

To provide further evidences, we shall now
present new lattice results. These suggest that
the discrepancy between the zero-momentum
$Z_\varphi$ and its perturbative prediction $Z_{\text{pert}}\sim Z_{\rm prop}$
is due
to the presence of a non-vanishing scalar condensate in the broken phase.
To this end, we have performed
finite-temperature simulations
by considering asymmetric
lattice, $L_s^3 \times L_t$,  periodic in time direction. As it is well
known, this is equivalent
to a non-zero temperature $T=1/L_t$.

At  $\kappa=0.07512$ (and for all values of $L_s$) there is clear
evidence for a phase transition in the region $6 < L_t < 8$ where
the system crosses from the broken into the symmetric phase (see
Fig.~1).
\begin{figure}[!ht]
\begin{center}
\includegraphics[width=0.4\textwidth,clip]{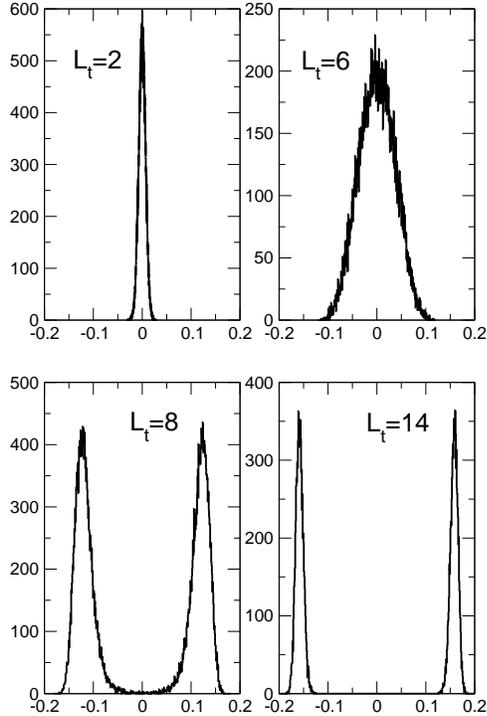}
\vspace{-1.0cm} \caption{The distribution of the average field for
$L_s=64$ and four values of $L_t$.}
\end{center}
\end{figure}
Our lattice results show that,
well above the phase transition temperature, i.e. for
very small $L_t$, when the system is in the symmetric phase, the data for the
propagator are well reproduced by Eq.~(\ref{gbetaform}) down to $p=
0$. Therefore,
at high temperature the zero-momentum limit is smooth
so that $Z_\varphi$ and $Z_{\text{prop}}$ agree to
very good accuracy, as they do in a $T=0$ symmetric-phase
simulation for $\kappa < \kappa_c$. Thus the broken-phase
discrepancy between $Z_\varphi$ and $Z_{\text{prop}}$ is a real
physical effect that could be ascribed to the presence of a non-vanishing
scalar condensate in the broken phase.
In Fig.~2 we have summarized our data for $Z_\varphi$.
They show a clear increase when $T \to T_c$ in contrast with the values of
$Z_{\mathrm{prop}}$ (not shown) that remain
remarkably close to the zero-temperature value $0.9551$. The trend for $T
\to T_c$ is in qualitative agreement with our previous zero-temperature
study for
$\kappa \to \kappa_c$.
\begin{figure}[!ht]
\begin{center}
\includegraphics[width=0.4\textwidth,clip]{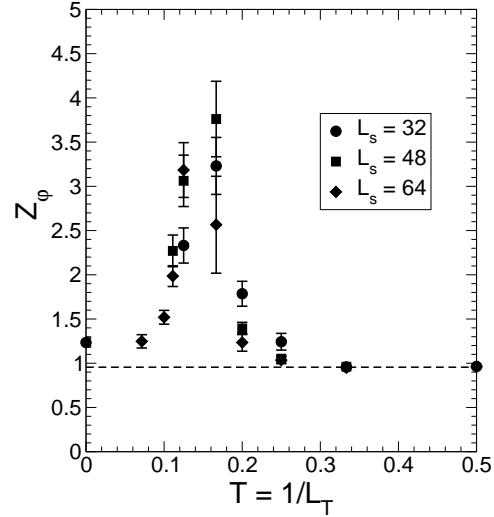}
\vspace{-1.0cm} \caption{Lattice data for $Z_\varphi \equiv
m^2_{\mathrm{prop}} \chi$ vs. T. Dashed line is the T=0 value
$Z_{\mathrm{prop}}=0.9551$.}
\end{center}
\end{figure}
We stress that the conventional interpretation of `triviality',
assuming $Z_\varphi = Z_{\mathrm{prop}} \simeq 1$, predicts that
approaching the phase transition one should find $m^2 \to 0$ and
$\chi \to \infty$ in such a way that $m^2 \chi$ remains constant.
In this sense, the finite-temperature simulations, showing a
dramatic increase of $Z_\varphi=2 \kappa\chi m^2_{\text{prop}}$
when approaching the phase transition, confirm and extend the
previous zero-temperature results where the continuum limit was
approached by letting $\kappa \to \kappa_c$. Both show that
$Z_\varphi$ is very different from the more conventional quantity
$Z_{\text{prop}} \sim 1$ determined perturbatively from the
residue of the shifted field propagator. For this reason, the
lattice data support once more the alternative picture of
Refs.~\cite{Consoli:1994jr,Consoli:1999ni} where an infinitesimal
curvature of the effective potential
$V''_{\text{eff}}=M^2_h/Z_\varphi \to 0$ can be reconciled with
finite values of $M_h$. This non-trivial result may have important
consequences for particle physics and cosmology


\begin{thebibliography}{1}

\bibitem{Consoli:1994jr}
M. Consoli and P.M. Stevenson,
\newblock Z. Phys. C63 (1994) 427, hep-ph/9310338.

\bibitem{Consoli:1999ni}
M. Consoli and P.M. Stevenson,
\newblock Int. J. Mod. Phys. A15 (2000) 133, hep-ph/9905427.

\bibitem{Cea:1999kn}
P. Cea et~al.,
\newblock Mod. Phys. Lett. A14 (1999) 1673, hep-lat/9902020.

\bibitem{Luscher:1988ek}
M. L{\"u}scher and P. Weisz,
\newblock Nucl. Phys. B295 (1988) 65.

\end{thebibliography}

\end{document}